\documentclass[%
 aip,
 jcp,
 amsmath,amssymb,
 reprint,%
]{revtex4-2}

\usepackage{graphicx}
\usepackage{dcolumn}
\usepackage{bm}

\usepackage[utf8]{inputenc}
\usepackage[T1]{fontenc}
\usepackage{mathptmx}
\usepackage{amsmath}
\usepackage{etoolbox}
\usepackage[hidelinks]{hyperref}
\usepackage{cancel}

\makeatletter
\def\@email#1#2{%
 \endgroup
 \patchcmd{\titleblock@produce}
  {\frontmatter@RRAPformat}
  {\frontmatter@RRAPformat{\produce@RRAP{*#1\href{mailto:#2}{#2}}}\frontmatter@RRAPformat}
  {}{}
}%
\makeatother
\begin{document}

\preprint{AIP/123-QED}

\title{Work statistics in slow thermodynamic processes}

\author{Jie Gu}
 \email{jiegu@ufl.edu}
 \affiliation{Chengdu Academy of Education Sciences, Chengdu 610036, China}

\date{\today}

\begin{abstract}
We apply the adiabatic approximation to slow but finite-time thermodynamic processes and obtain the full counting statistics of work.
The average work consists of  change in free energy and the dissipated work, and we identify each term as a dynamical- and geometric-phase-like quantity. 
An expression for the friction tensor, the key quantity in thermodynamic geometry, is explicitly given.
The dynamical and geometric phases are proved to be related to each other via the fluctuation-dissipation relation. 
\end{abstract}

\maketitle

\section{Introduction}
\label{sect:intro}

Reversibility is valid only for infinitely long quasi-static transformations, in which both dissipation and fluctuations are absent. 
Despite perfect efficiency and constancy, these processes are not realistic.
Slow but finite-time processes add non-trivial ingredients in quasi-static processes, while remaining analytically tractable.
In slow thermodynamic processes, the driving forces are varying slowly compared with relaxation time scales. It implies that after an exponentially short transient depending on the initial states, the state is all the time near the instantaneous steady state  \cite{Cavina2017}.

There are at least five frameworks for investigating slow thermodynamic processes. 
The first framework is a perturbation theory developed in Refs. \cite{Mandal2016,Cavina2017}. 
The slowness of driving allows a perturbative expansion of the   probability distribution or density matrix in powers of $\tau^{-1}$, where $\tau$ is the time duration of the process. 
The second framework is the adiabatic response theory \cite{Sivak2012,Zulkowski2012}, which is an extension of the well-established dynamic linear response theory \cite{zwanzig2001nonequilibrium}. It exploits the property that states are close to equilibrium.  
Both frameworks reveal a geometric structure of thermodynamic processes in small systems \cite{Crooks2007,Zulkowski2012,Zulkowski2015,Miller2019,Scandi2019,Scandi2020,Miller2020,Abiuso2020,  Brandner2020,Frim2022}, extending the concept of thermodynamic length for macroscopic endoreversible thermodynamics \cite{Kirkwood1946,Weinhold1975,Ruppeiner1979,Salamon1983,Janyszek1989,Brody1995,Ruppeiner1995}. The key notion is the so-called \emph{friction tensor} \cite{Kirkwood1946,Sivak2012,Zulkowski2012, Mandal2016} (also called \emph{thermodynamic metric tensor} \cite{Miller2020, Abiuso2020,Frim2022}). The dissipated work during $[0,\tau]$ is expressed in terms of the friction tensor $\boldsymbol{g}$ as
\begin{equation}
    W_\text{diss} = \int_0^\tau \dot{\boldsymbol{\Lambda}}\cdot \boldsymbol{g} \cdot \dot{\boldsymbol{\Lambda}}^\text{T}   \text{d}t,
\end{equation}
where $W_\text{diss}=\langle W\rangle -\Delta F$, i.e., the dissipated work $W_\text{diss}$ is the difference
between average work $\langle W\rangle$ done on the system and the change in equilibrium free energy $\Delta F$, and $\boldsymbol{\Lambda}$ denotes time-dependent control parameters. By the Cauchy-Schwarz inequality, the dissipated work satisfies
\begin{equation}
\label{eq:thermodynamiclength}
    \tau W_\text{diss} \ge \mathcal{L}^2,
\end{equation}
where $\mathcal{L} =\int_0^\tau \sqrt{\dot{\boldsymbol{\Lambda}}\cdot \boldsymbol{g} \cdot \dot{\boldsymbol{\Lambda}}^\text{T} }  \text{d}t $ is the thermodynamic length as defined by the metric.

The third framework treats a slow process as a sequence of  quench-and-relaxation processes where the number of steps is large but finite.
Its equivalence with the previous two frameworks has been proved in Ref. \cite{Scandi2020}.
The fourth is stochastic thermodynamics \cite{Seifert2008,Seifert2012}. As an nonperturbative formalism, its applicability is not limited to slow processes. 
For a system in contact with a heat bath with inverse temperature $\beta$ and undergoing slow processes, the work distribution is Gaussian, and the
Jarzynski equality \cite{Jarzynski1997a,Jarzynski1997}, one of the most notable results in this framework, gives the fluctuation-dissipation
relation (FDR)
\begin{equation}
\label{eq:fdr}
    W_\text{diss} = \frac{1}{2}\beta \sigma_W^2,
\end{equation}
where the dissipated work is related to the variance of work $\sigma_W^2$. The FDR in this form has also been derived by the third framework \cite{Scandi2020} and  a variant of the perturbation theory (the first framework) \cite{Speck2004}.

The fifth framework applies the adiabatic approximation in quantum mechanics to classical stochastic processes  for the full counting statistics of thermodynamic quantities \cite{Sinitsyn2007,Sinitsyn2009}.
It was introduced for classical chemical kinetics \cite{Sinitsyn2007}, but has been soon extended to a geometrical expression of excess entropy production \cite{Sagawa2011}, particle statistics in quantum transport problems\cite{Ren2010, Yuge2012,Goswami2016,Gu2017} and quantum heat engines \cite{Giri2017}, in which Berry-phase-induced effects are the focus \cite{Berry1984}.
Nonetheless, this simple but powerful framework has not been applied to work statistics to the best of our knowledge.
In addition, the connections between the last framework and the other four have not been explored. Both the Berry's geometric phase  and the thermodynamic metric tensor are geometric  quantities. Is there any connection between them?
Since the fifth framework enables one to calculate the full counting statistics of work, the variance of work can in principle be obtained, so what role does the fluctuation-dissipation relation play in this framework? In this study, we will answer these two questions and unveil the connections among different frameworks.

\section{Full Counting Statistics of Work}
Consider a system with $N$ energy levels $\{\epsilon_m(t)\}$ interacting with a heat bath with fixed inverse temperature $\beta$. 
The energy levels $\{\epsilon_m(t)\}$ are varying slowly by external manipulations, serving as control parameters of the protocol.
Assume the system dynamics follows a stochastic process modeled by a continuous-time Markov chain. 
The time evolution of the probability distribution
\({{|p(t)\rangle}}=[p_1,p_2,\ldots, p_N]^\text{T}\) is described by a Pauli master equation \cite{van1992stochastic,breuer2002theory}
\begin{equation}
    \frac{\text{d} |{p}(t)\rangle }{\text{d} t}= \boldsymbol{R}(t) |p(t)\rangle,
\end{equation}
where \(p_m\), the $m$-th entry of $|p\rangle$, is the probability of state \(m\); \(R_{mn} \ (n \ne m)\),  the $mn$-th entry of the matrix $\boldsymbol{R}$, is the time-dependent transition rate from state \(n\) to \(m\), and
\(R_{mm} = -\sum_{n,n\ne m} R_{nm}\).
We assume that the Markov chain is ergodic and the detailed balance condition
\begin{equation}
\label{eq:dbalance}
    \frac{R_{mn}}{R_{nm}} = e^{\beta (\epsilon_n-\epsilon_m)}
    \end{equation}
is satisfied, so at any instant there exists an instantaneous Gibbs state $|\lambda (t)\rangle$ satisfying $\boldsymbol{R}(t)|\lambda (t)\rangle = \boldsymbol{0}$.

The full counting statistics of work can be obtained as follows. Let \(P(W,\tau)\) be the probability of having done work \(W\) on the
system within time \([0,\tau]\). 
The full counting statistics of work
is characterized by the cumulant generating function \(\mathcal{K}_\chi(\tau)\) and moment generating function \(\mathcal{M}_\chi(\tau)\) \cite{van1992stochastic},
\begin{equation}
  \mathcal{K}_\chi(\tau) = \ln \mathcal{M}_\chi(\tau) \equiv  \ln \int_{-\infty}^\infty e^{i\chi W}P(W,\tau) \text{d}W,  
\end{equation}
where \(\chi\) is the counting field. 
Then the average and variance of
work are given by, respectively,
\begin{equation}
  \langle W \rangle = \mathcal{K}' \equiv \left.\frac{\partial \mathcal{K}_\chi(\tau)}{\partial (i\chi)} \right|_{\chi = 0}  
\end{equation}
and
\begin{equation}
  \sigma^2_W = \mathcal{K}'' \equiv \left.\frac{\partial^2 \mathcal{K}_\chi(\tau)}{\partial (i\chi)^2} \right|_{\chi = 0}.  
\end{equation}

In order to find the cumulant generating function, let
\(P_{m}\left(w, t\right)\) denote the probability that within time \(t\)
the work done to system is \(w\) and the system is dwelling on the
\(m\)-th state. Then \(P_{m}\left(w+d w, t+d t\right)\) consists of two
contributions given by
\begin{equation}
    \begin{aligned}
    &P_{m}\left(w+\text{d} w, t+\text{d} t\right) \\
    &=P_{m}\left(w, t\right)\left(1+R_{mm} \text{d} t\right)+\sum_{n(\ne m)} P_{n}\left(w+\text{d} w, t\right) R_{mn} \text{d} t.
    \end{aligned}
\end{equation}
The first term represents that the system dwell on the \( m  \)-th
state during \([t,t+\text{d} t)\) , and the energy level $m$-th state is
manipulated so the work $\text{d}w=\text{d}\epsilon_m$ is performed on the system on top of $w$. The
second term represents jumps from other states, during which no work is
done. Up to the first order of $\text{d}t$, the equation becomes
\begin{equation}
  \frac{\partial P_{m}}{\partial w} \dot{\epsilon}_{m}+\frac{\partial P_{m}}{\partial t}=P_{m} R_{mm}+\sum_{n(\ne m)} P_{n} R_{mn}  
\end{equation}

By multiplying \(e^{i\chi w}\) on both sides and integrating with
respect to \(w\), the time evolution of the state-resolved moment
generating function is given by
\begin{equation}
\label{eq:diff}
  \frac{\partial \mathcal{M}_{\chi,m}(t)}{\partial t}=\sum_{n} {R}_{\chi,mn}(t) \mathcal{M}_{\chi,n}(t)  
\end{equation}
where
\begin{equation}
  {R}_{\chi,mn}(t) = R_{mn}(t) +i \chi \dot{\epsilon}_{m}(t)\delta_{mn},  
\end{equation}
or in matrix form,
\begin{equation}
\label{eq:Rchi}
  \boldsymbol{R}_\chi(t) = \boldsymbol{R}(t)+i\chi \dot {\boldsymbol{H}}(t),
\end{equation}
where the Hamiltonian $  \boldsymbol{H}(t)=\text{diag}\{\epsilon_m(t)\}.  $
Similar results for a two-state system has been obtained in Ref. \cite{Verley2013}.


\section{Main results}

For later use, we define the eigenvalue of \(\boldsymbol R_\chi(t)\)
with the largest real part as \(\lambda_\chi(t)\), with the
corresponding left and right eigenvectors being
\( \langle \lambda_\chi(t)|  \) and \(|\lambda_\chi(t)\rangle\) by a slight abuse of notation.
The left and right eigenvectors associated with same eigenvalue are not Hermitian-conjugate of each other due to the non-Hermiticity of $\boldsymbol{R}_\chi$.
Evaluating at \(\chi=0\) gives \(\lambda_{\chi=0}=0\),
\(\langle \lambda| \equiv \langle \lambda_{\chi=0}(t)|  = [1,1,\ldots,1]\)
and
\(|\lambda(t) \rangle \equiv |\lambda_{\chi=0}(t)\rangle =|p^\text{G}(t)\rangle\),
the instantaneous Gibbs state, i.e.,
\begin{equation}
\label{eq:gibbs}
    |\lambda (t) \rangle = \frac{1}{Z} \big[e^{-\beta \epsilon_1(t)} , e^{-\beta \epsilon_2(t)},\ldots , e^{-\beta \epsilon_N(t)} \big]^\text{T},
\end{equation}
where \(Z = \text{Tr} [\exp(-\beta \boldsymbol{H})]\) is the partition function.

The cumulant generating function is then
\begin{equation}
  \mathcal{K}_\chi(\tau) =\ln\sum_m \mathcal{M}_{\chi,m}(\tau) = \ln \langle \lambda| \mathcal{M}_\chi(\tau) \rangle,  
\end{equation}
where we have defined
\(| \mathcal{M}_\chi(\tau)\rangle = [\mathcal{M}_{\chi,1},\mathcal{M}_{\chi,2},\ldots,\mathcal{M}_{\chi,N}]^\text{T}\).

Given the differential equations Eq. \eqref{eq:diff}, the cumulant generating function
can be formally solved by
\(\mathcal{K}_\chi (\tau) = \ln \langle \lambda |\mathbb{T}\big[\exp({\int_{0}^{\tau} \boldsymbol{R}_\chi( t) d t}\big] |p(0)\rangle\),
where \(\mathbb{T}\) is the time-ordering operator, and 
\(|p(0)\rangle\) is the initial state. We consider slow processes
and assume that the initial state is the Gibbs state
\( |p(0)\rangle = |\lambda(0)\rangle  \). 
Given the similarity between Eq. \eqref{eq:diff} and the Schr\"odinger equation, by invoking the adiabatic
approximation analogous to that in quantum mechanics and following Refs.
\cite{Sinitsyn2007,Sinitsyn2009,Ren2010,Sagawa2011, Gu2017}, we find
\begin{equation}
  \mathcal{M}_\chi(\tau)=e^{\int_{0}^{\tau} \lambda_{\chi}(t) \text{d} t}e^{-\int_{C}\left\langle \lambda_{\chi}|\dot \lambda_{\chi}\right\rangle \text{d}t} \left\langle \lambda_{\chi}(0) | \lambda(0)\right\rangle \left\langle\lambda|\lambda_\chi(\tau)\right\rangle  
\end{equation}
or equivalently
\begin{subequations}
\begin{align}
  &
  \label{eq:Kchi}
  \mathcal{K}_\chi=\mathcal{K}_\text{dyn}+\mathcal{K}_\text{geo}+\big [\ln\left\langle \lambda_{\chi}(0) | \lambda(0)\right\rangle +\ln \left\langle\lambda|\lambda_\chi(\tau)\right\rangle \big ],   \\
  & \mathcal{K}_\text{dyn}=\int_{0}^{\tau} \lambda_{\chi}(t) \text{d} t,  \\
\label{eq:decomposition}
& \mathcal{K}_\text{geo}=-\int_0^\tau \langle \lambda_{\chi}|\dot \lambda_{\chi}\rangle \text{d}t.
\end{align}
  \end{subequations}
We have omitted the argument \(\tau\) for convenience. It is clear that
\(\mathcal{K}_\text{dyn}\) and \(\mathcal{K}_\text{geo}\) are analogous to the dynamical and
Berry's geometric  phase in quantum mechanics, respectively \cite{Berry1984}. The dynamical-phase part corresponds to a fictitious process during which each state at an arbitrary instant is the instantaneous Gibbs state, and the geometric-phase part corresponds to effects on top of the former due to the slow but finite-time driving.
As will be seen shortly, $\mathcal{K}_\text{geo}$ here is not a truly geometric quantity, but we will still keep the subscript ``geo'' due to its apparent resemblance to the geometric phase in quantum mechanics.
The boundary terms (the two terms in the square bracket) are reminiscent of the noncyclic geometric phase \cite{Ohkubo2010}.

Now we proceed
to find the average and variance of work by taking  the first and second
derivatives of \(\mathcal{K}_\chi\). 
Differentiating both sides of
\(\langle \lambda_\chi|  \boldsymbol{R}_\chi = \lambda_\chi \langle \lambda_\chi|\)
gives
\begin{equation}
\label{eq:firstderivative}
  \langle \lambda_\chi'| \boldsymbol{R}_\chi + \langle \lambda_\chi| \boldsymbol{R}_\chi' = \lambda_\chi'\langle \lambda_\chi|+\lambda_\chi\langle \lambda_\chi'|  ,
\end{equation}
where we have defined \(A_\chi'\equiv\partial A_\chi/\partial(i\chi)\) and $A'\equiv\partial A_\chi/\partial(i\chi) |_{\chi=0}$.

By evaluating Eq. \eqref{eq:firstderivative} at \(\chi=0\) and multiplying both sides by \(|\lambda \rangle\) we
obtain the derivative of the eigenvalue,
\begin{equation}
\label{eq:lambdaprime}
  \lambda' = \langle \lambda| \dot{\boldsymbol{H}} | \lambda\rangle=\text{Tr}(\dot{\boldsymbol{H}} e^{-\beta \boldsymbol{H} })/ Z,  
\end{equation}
This result obviously coincides with the time-independent first-order perturbation theory.

By evaluating Eq. \eqref{eq:firstderivative} at \(\chi=0\) and multiplying by the Drazin inverse \(\boldsymbol{R}^+\)
we obtain
\begin{equation}
\label{eq:left}
  \langle \lambda' | = - \langle \lambda|\dot{\boldsymbol{H}} \boldsymbol{R}^+,  
\end{equation}
where we have used the property of Drazin inverse that
\(\langle \lambda|\boldsymbol{R}^+ = \boldsymbol{0}\) \cite{Mandal2016,Crooks2018}, the definition of
\(\boldsymbol{R}_\chi\) and the fact that \(\lambda_{\chi=0}=0\). See Appendix \ref{sect:appendix} for definition and properties of Drazin inverse. 

The derivative of the right eigenvector can be obtained similarly,
\begin{equation}
\label{eq:right}
  | \lambda' \rangle = -\boldsymbol{R}^+ \dot{\boldsymbol{H}} | \lambda \rangle.  
\end{equation}

According to Eq. \eqref{eq:lambdaprime}, the derivative of the dynamical phase is, up to the
first order,
\begin{equation}
  \mathcal{K}'_\text{dyn} = \int_0^\tau \lambda' dt = -\beta^{-1}\ln \frac{Z(\tau)}{Z(0)} .  
\end{equation}
Therefore, we obtain the identity
\begin{equation}
\label{eq:Kprimedyn}
\Delta F = \mathcal{K}'_\text{dyn},
\end{equation}
where \(F(t)=-\beta^{-1}\ln Z(t)\) is the equilibrium free energy.
This makes sense because as mentioned above, the dynamical-phase part corresponds to a fictitious process during which the system is in equilibrium at any instant, and for this kind of process, the average work is equal to the change in equilibrium free energy.

By using the properties of Drazin inverse again, we prove in Appendix \ref{sect:appendix} that the boundary terms do not contribute to $\mathcal{K}'$, so \(\langle W \rangle = \mathcal{K}'=\mathcal{K}'_\text{dyn}+\mathcal{K}'_\text{geo}\).
Since \(\langle W\rangle = \Delta F+W_\text{{diss}}\), Eq. \eqref{eq:Kprimedyn} implies that the dissipated
work \(W_\text{diss}\) can be identified as the derivative of the geometric phase,
\begin{equation}
\label{eq:Kprimegeo}
W_\text{diss}= \mathcal{K}'_\text{geo}.
\end{equation}

Plugging Eqs. \eqref{eq:left} and \eqref{eq:right} into Eq. \eqref{eq:decomposition} gives the dissipated work:
\begin{equation}
\label{eq:geo}
\begin{aligned}
 W_\text{diss}= \mathcal{K}'_\text{geo} &= \int_0^\tau \langle \lambda|\dot{\boldsymbol{H}} \boldsymbol{R}^{+} |\dot \lambda\rangle \text{d}t \\
  & =-\beta \int_0^\tau \langle \lambda|\dot{\boldsymbol{H}} \boldsymbol{R}^+ \dot{\boldsymbol{H}} |\lambda\rangle \text{d}t ,
\end{aligned}
  \end{equation}
where in order to obtain the second line we have used Eq. \eqref{eq:gibbs} and the property of Drazin inverse that \(\boldsymbol{R}^+ |\lambda \rangle =\mathbf{0}\).
One can see the symmetric structure in Eq. \eqref{eq:geo}. 
The counterpart of Eq. \eqref{eq:geo} in open quantum systems has been derived in Ref. \cite{Scandi2019} by the perturbation theory \cite{Cavina2017}, and reduces to our result for classical states, i.e., when the initial state is a Gibbs state, and $[H(t),H(t')]=0$ such that there is no drive-induced coherence throughout.

As derived in Appendix \ref{sect:appendix}, the dissipated work can be explicitly written as
\begin{equation}
\label{eq:wdiss}
  W_\text{diss}=\mathcal{K}_\text{geo}' =  \int_0^\tau \sum_{m,n}\dot \epsilon_m g_{mn} \dot \epsilon_n \text{d}t,  
\end{equation}
where the {friction tensor} ({thermodynamic metric tensor}) is
\begin{equation}
\label{eq:tensor}
  g_{mn} = - \frac{\beta \big[R_{mn}^+ e^{-\beta \epsilon_n}+R_{nm}^+ e^{-\beta \epsilon_m}\big]}{2Z}.  
\end{equation}
Here, $R_{mn}^+$ is the $mn$-th entry of $\boldsymbol{R}^+$.
This tensor is symmetric and positive semi-definite.
The explicit expression of the friction tensor for slow classical thermodynamic processes has not been reported to our knowledge.
From Eqs. \eqref{eq:wdiss} and \eqref{eq:tensor}, it can be seen that the dissipated power $\dot W_\text{diss}$ and entropy production rate in slow processes is time-local: they only depends on the instantaneous values and derivatives of the control parameters \cite{Sivak2012}.

Now let us prove the fluctuation-dissipation theorem for this setup.
Differentiating both sides of
\(\langle \lambda_\chi|  \boldsymbol{R}_\chi = \lambda_\chi \langle \lambda_\chi|\)
twice gives
\begin{equation}
\label{eq:lambdappleft}
  \langle \lambda_\chi''| \boldsymbol{R}_\chi +2\langle \lambda_\chi'| \boldsymbol{R}_\chi'  = \lambda_\chi''\langle \lambda_\chi|+2\lambda_\chi'\langle \lambda_\chi'|+\lambda_\chi\langle \lambda_\chi''|.  
\end{equation}
Evaluating at \(\chi=0\) and multiplying by \(|\lambda \rangle\) lead to
\begin{equation}
  \lambda'' =  -2 \langle \lambda|\dot{\boldsymbol{H}} \boldsymbol{R}^+ \dot{\boldsymbol{H}} |\lambda\rangle.  
\end{equation}

In Appendix \ref{sect:appendix} we show that, up to the first order of $\tau^{-1}$, the second derivatives of terms other than the dynamical phase do not contribute. Thus,
\begin{equation}
 \sigma_W^2 = \mathcal{K}''= \mathcal{K}_\text{dyn}'' =    \int_0^\tau \lambda''\text{d}t = -2  \int_0^\tau \langle \lambda|\dot{\boldsymbol{H}} \boldsymbol{R}^+ \dot{\boldsymbol{H}} |\lambda\rangle \text{d}t.
\end{equation}
Comparing this equation with Eq. \eqref{eq:geo}, we immediately obtain the fluctuation-dissipation relation in the form of Eq. \eqref{eq:fdr} \cite{Speck2004,Miller2019,Scandi2019,Scandi2020}. Note that the geometric phase does not vanish in general and still contributes, but it is a higher order term of $\tau^{-1}$ compared with the dynamical phase.

Next we discuss the condition under which the dissipated work and also the fluctuation of work vanish.
Obviously the dissipated work is vanishingly small for a quasi-static (infinitely slow) process.
Other than the previous trivial case, it is not hard to see that there is no fluctuation of work if 
all energy levels are shifted globally; the populations are unchanged, so there is no entropy production associated with this process.
Mathematically, this condition implies that the rates of change in energy levels are identical, i.e., $ \dot{\boldsymbol{H}} \propto \boldsymbol{I} $, where $\boldsymbol{I}$ is the identity operator. Plugging it into 
Eq. \eqref{eq:geo}, we indeed have $\sigma_W^2 \propto W_\text{diss} = 0$, due to the property that $\langle \lambda | \boldsymbol{R}^+ =\mathbf{0}$ \cite{Mandal2016,Crooks2018}.
To prove the necessity of the condition, we exploit the strong result by Shiraishi \emph{et al.} \cite{shiraishi2018},
\begin{equation}
    \Sigma \ge \frac{L(|p(0)\rangle, |p(\tau)\rangle )^2}{2A \tau},
\end{equation}
where $\Sigma$ is the entropy production, $\tau$ the time duration of the process, $L(|p(0)\rangle, |p(\tau)\rangle )$ the trace distance between $|p(0)\rangle$ and $ |p(\tau)\rangle $, and $A$ the dynamical activity.
This speed limit implies that, in order for zero entropy production during any finite-time processes, the probability distribution does not change with time.
Combining with the detailed balance condition, one can see that the entropy production vanishes only if the energy level spacing between an arbitrary pair of energy levels remains constant.
Furthermore, it remains an open question whether vanishing entropy is a sufficient or necessary condition (or neither) for the fluctuation relation for currents to hold \cite{sinitsyn2011}, an interesting connection we will not pursue any further here.

On the other hand, one might recall that, in quantum mechanics, if there is only one control parameter in the Hamiltonian that is changing with time periodically, i.e., $\boldsymbol{H}(\Lambda(t))=\boldsymbol{H}(\Lambda(t+\mathcal{T}))$ where $\mathcal{T}$ is the period, the geometric phase vanishes for this case \cite{Griffiths2004}. It is also well known that no net charges are transported in a single-parameter adiabatic pump  \cite{Switkes1999,Ren2010, Yuge2012,Goswami2016,Gu2017,Giri2017}; during the second half period, any charge that has flowed during the first will return.
Therefore, one might expect $W_\text{diss}=\mathcal{K}_\text{geo}'=0$ if there is only one control parameter.
Nonetheless, similar reasoning does not apply to $W_\text{diss}$ because the dissipated work and entropy production are 
monotonically increasing with time; the integrand in Eq. \eqref{eq:wdiss} is not guaranteed to vanish and in general positive even if there is only one control parameter (see the next section for a concrete example), so the dissipated work during the second half period obviously does not cancel the first half.
In other words, $W_\text{diss}$ is generally positive even for single-parameter periodic systems,  contradicting the expectation above.
This contradiction can be resolved by the observation that $\mathcal{K}_\text{geo}$ is not truly geometric even if Eq. \eqref{eq:decomposition} shows a  structure similar to the geometric phase.
A geometric quantity should be independent of how the path in the parameter space is traversed, especially of the changing rate of control parameters (of course as long as the adiabatic approximation holds).
However, $\langle \lambda_\chi|$ and $|\lambda_\chi \rangle$ depend on not only the control parameters $\boldsymbol{\Lambda}(t)$, but also the time derivative ${\dot{\boldsymbol{\Lambda}}}(t)$ as can be seen from Eq. \eqref{eq:Rchi}.
Therefore, $\mathcal{K}_\text{geo}$ and $W_\text{diss}$ are not geometric quantities. By contrast, the geometric phase in quantum mechanics \cite{Berry1984}, geometric-phase-induced transport \cite{Switkes1999,Ren2010, Yuge2012,Goswami2016,Gu2017,Giri2017} and the thermodynamic length defined below Eq. \eqref{eq:thermodynamiclength} are genuinely geometric.

\section{Example}
As a simple illustration, let us consider a two-state system. The transition rate matrix is given by
\begin{equation}
  \boldsymbol{R} = \begin{bmatrix} -k_1 & k_2 \\ k_1 & -k_2 \end{bmatrix}.
\end{equation}
The eigensystem associated to the nonzero eigenvalue is 
\begin{equation}
\begin{aligned}
  &\lambda_{1} =-\left(k_{1}+k_{2}\right), \\
 & |\lambda_1 \rangle = \begin{bmatrix} 1, & -1 \end{bmatrix}^\text{T},\\
 & \langle \lambda_1| = \frac{1}{k_1+k_2}\begin{bmatrix} k_1, & -k_2 \end{bmatrix},
\end{aligned}
\end{equation}
so the Drazin inverse is simply given by
\begin{equation}
  \boldsymbol{R}^+ = \frac{1}{(k_1+k_2)^2}\boldsymbol{R}.  
\end{equation}

By using the first-order perturbation theory, we find, up to the first order of $\chi$, the eigensystem of $\boldsymbol{R}_\chi$ associated to $\lambda_\chi$ (the eigenvalue with the largest real part) is:
\begin{subequations}
    \begin{align}   
    \lambda_\chi & = \frac{i\chi}{2} \left[ (\dot \epsilon_1+\dot \epsilon_2) + \frac{(k_1-k_2)(\dot \epsilon_2 - \dot \epsilon_1)}{k_1+k_2}    \right], \\
    \label{eq:right2}
    |\lambda_\chi \rangle & = \frac{1}{k_1+k_2} \begin{bmatrix} k_2, & k_1 \end{bmatrix}^\text{T} + \frac{i\chi k_1 k_2(\dot \epsilon_1 \dot \epsilon_2)}{(k_1+k_2)^3} \begin{bmatrix} 1,&-1\end{bmatrix}^\text{T}, \\
    \label{eq:left2}
    \langle \lambda_\chi| &= \begin{bmatrix} 1, & 1 \end{bmatrix} + \frac{i\chi (\dot \epsilon_1 - \dot \epsilon_2)}{(k_1+k_2)^2} \begin{bmatrix}k_1,&-k_2 \end{bmatrix}.
\end{align}
\end{subequations}
Exploiting the detailed balance condition Eq. \eqref{eq:dbalance}, we have
\begin{equation}
    \lambda' = \left. \frac{\partial \lambda_\chi}{\partial (i\chi)} \right|_{\chi=0} = \frac{e^{-\beta \epsilon_1} \epsilon_1 + e^{-\beta \epsilon_2} \epsilon_2}{Z} 
\end{equation}
The R.H.S. coincides with $\dot F$, the time derivative of equilibrium free energy, so we have verified Eq. \eqref{eq:Kprimedyn}, i.e., the dynamical-phase part of $\mathcal{K}'$ is equal to the change in equilibrium free energy.

Substituting Eqs. \eqref{eq:right2} and \eqref{eq:left2} into Eq. \eqref{eq:decomposition} and using the detailed balance condition, one can obtain the dissipated power
\begin{equation}
\label{eq:wdiss_twostate}
  \dot W_\text{diss}= \frac{\beta(\dot \epsilon_1-\dot \epsilon_2)^2 k_1 k_2}{(k_1+k_2)^3}. 
\end{equation}
It is straightforward to check that plugging the Drazin inverse into Eq. \eqref{eq:geo} results in exactly the same result.
It verifies the discussion in the previous section that the dissipated power vanishes if and only if $\dot \epsilon_1=\dot \epsilon_2$.
Accordingly, the friction tensor is
\begin{equation}
    \boldsymbol{g} =\frac{\beta k_1 k_2}{(k_1+k_2)^3} \begin{bmatrix}
     1& -1 \\
     -1 & 1
    \end{bmatrix}.
\end{equation}
This tensor is positive semi-definite as expected.

Even more concretely, we consider a special two-state system---a model for the bit erasure \cite{Landauer1961}, whose Hamiltonian is given by
\begin{equation}
  \boldsymbol{H}= \frac{\Lambda(t)}{2}\sigma_z.  
\end{equation}
During the population reduction step, the energy level splitting $\Lambda(t)$ is increased from zero to an energy much greater than $\beta^{-1}$.
The transition rates are related to $\Lambda(t)$ as
\begin{equation}
  k_1 = \gamma n(\Lambda),\quad k_2 = \gamma [n(\Lambda)+1], 
\end{equation}
where $\gamma$ is the coupling between the system and bath, and \(n(\Lambda)=(e^{\beta \Lambda}-1)^{{-1}}\) is the Bose-Einstein distribution function. Plugging the transition rates into Eq. \eqref{eq:wdiss_twostate} leads to
\begin{equation}
  \dot{W}_{\mathrm{diss}}=\frac{\beta\dot{\Lambda}^{2}\left(1-e^{-\beta \Lambda}\right) e^{-\beta \Lambda}}{\gamma\left(1+e^{-\beta \Lambda}\right)^{3}}.  
\end{equation}
Even though there is only one control parameter, $W_\text{diss}$ remains positive.
We note that exactly the same expression has been derived recently by the perturbation theory \cite{Ma2021}. 

\section{Conclusion}
The adiabatic approximation is a powerful tool not only in quantum mechanics, but also in slow stochastic processes. 
With this tool, the following connections have been unveiled.
From Eqs. \eqref{eq:Kprimedyn} and \eqref{eq:Kprimegeo}, one can see that $\langle W \rangle$ consists of $\Delta F$ and $W_\text{diss}$, with each contribution identified as a dynamical- and geometric-phase-like quantity, respectively. 
Furthermore, in Eqs. \eqref{eq:wdiss} and \eqref{eq:tensor}, we calculated the friction tensor $\boldsymbol{g}$ from the geometric phase, which is consistent with the results obtained by the first two frameworks described in Sect. \ref{sect:intro}. The fluctuation-dissipation relation was derived in an elementary and transparent manner, complementary to the third and fourth frameworks. 
These results can be summarized in a relation between the dynamical phase and geometric phase,
\begin{equation}
   \mathcal{K}'_\text{geo} = \frac{\beta}{2} \mathcal{K}''_\text{dyn}.
\end{equation}
Whether similar relations hold for higher cumulants and/or for other thermodynamic quantities is an interesting open question.
Results in slow quantum processes can be obtained by applying this framework to quantum master equations  \cite{Yuge2012,Goswami2016,Giri2017}, and can be compared with existent results such as Ref. \cite{Miller2019,Scandi2019,Scandi2020,Miller2020,Abiuso2020}.

\section*{Data Availability Statement}
The data that supports the findings of this study are available within the article.

\appendix
\section{}
\label{sect:appendix}
The Drazin inverse of a transition rate matrix $\boldsymbol{R}$ can be found as follows \cite{Mandal2016,Crooks2018}. Since  $\boldsymbol{R}$ is assumed to be diagonalizable, it can be expanded as
\begin{equation}
    \boldsymbol{R} = \sum_{\xi_\alpha \ne 0}\xi_\alpha |\xi_\alpha \rangle \langle \xi_\alpha |,
\end{equation}
where $|\xi_\alpha \rangle$ and $\langle \xi_\alpha |$ are, respectively, the left and right eigenvector of $\boldsymbol{R}$ associated with the eigenvalue $\xi_\alpha$. Together with $|\lambda\rangle$ and $\langle \lambda |$ that are associated with the zero eigenvalue, they satisfy \cite{brody2014}:
\begin{enumerate}
    \item $\langle \lambda|\xi_\alpha \rangle =0$, $\langle \xi_\alpha|\lambda \rangle=0$, $\langle \xi_\alpha|\xi_\beta \rangle=0$.
    \item $\langle \xi_\alpha | \xi_\alpha \rangle=1$, $\langle \lambda|\lambda \rangle=1$.
    \item $\sum_{\xi_\alpha \ne 0} |\xi_\alpha\rangle \langle \xi_\alpha | + |\lambda \rangle \langle \lambda| = \boldsymbol{I}$,     where $\boldsymbol{I}$ is the identity matrix.
\end{enumerate}

Its Drazin inverse is then
given by
\begin{equation}
    \boldsymbol{R}^+ = \sum_{\xi_\alpha \ne 0} \frac{1}{\xi_\alpha} |\xi_\alpha \rangle \langle \xi_\alpha |.
\end{equation}

The Drazin inverse has the following properties \cite{Mandal2016,Crooks2018}:
\begin{enumerate}
    \item \(\langle \lambda|\boldsymbol{R}^+ = \boldsymbol{0}\) because $\langle \lambda|\xi_\alpha \rangle =0$.
    \item Similarly, \(\boldsymbol{R}^+ |\lambda \rangle =\mathbf{0}\).
    \item The biorthogonality and completeness relations between eigenvectors imply
    \begin{equation}
        \begin{aligned}
\boldsymbol{R}^+ \boldsymbol{R} &= \sum_{\xi_\alpha \ne 0,\xi_\beta\ne 0} \frac{\xi_\beta}{\xi_\alpha} |\xi_\alpha \rangle \langle \xi_\alpha | \xi_\beta \rangle \langle \xi_\beta | \\
& = \sum_{\xi_\alpha \ne 0}  |\xi_\alpha \rangle \langle \xi_\alpha | \\
& = \boldsymbol{I}- |\lambda\rangle \langle\lambda|.
\end{aligned}
    \end{equation}
\end{enumerate}

Now we prove that the boundary terms do not contribute to $\mathcal{K}'$. The derivative of the boundary terms with respect to $\chi$ is
\begin{equation}
    \begin{aligned}
 & \left.   \frac{\partial \big [\ln\left\langle \lambda_{\chi}(0) | \lambda(0)\right\rangle +\ln \left\langle\lambda|\lambda_\chi(\tau)\right\rangle \big ]}{\partial \chi}
\right|_{\chi=0} \\
    =&  \left\langle \lambda'(0) | \lambda(0)\right\rangle +\left\langle\lambda|\lambda'(\tau)\right\rangle
    \end{aligned} 
\end{equation}
By substituting Eq. \eqref{eq:left} into $\left\langle \lambda'(0) | \lambda(0)\right\rangle $ and making use of  \(\boldsymbol{R}^+ |\lambda \rangle =\mathbf{0}\), we see that the first term vanishes. Similarly, the second term  also vanishes.

The explicit expression for the dissipated work, Eq. \eqref{eq:wdiss}, can be derived by expressing the vectors and matrices in terms of entries,
\begin{equation}
\langle \lambda|\dot{\boldsymbol{H}} \boldsymbol{R}^{+} |\dot \lambda\rangle
 = \sum_{m,n} \dot \epsilon_m R^+_{mn} \big( -\frac{\beta e^{-\beta \epsilon_n} \dot \epsilon_n}{Z}+\frac{\beta e^{-\beta \epsilon_n} }{Z^2} \sum \limits_l e^{-\beta \epsilon_l} \dot \epsilon_l\big).
\end{equation}
The second term vanishes due to the second property of Drazin inverse.
By symmetrizing the first term and plugging it into Eq. \eqref{eq:geo}, we obtain Eqs. \eqref{eq:wdiss} and \eqref{eq:tensor}.

Finally, let us prove that, up to the first order of $\tau^{-1}$, the second derivatives of terms other than the dynamical phase do not contribute to $\mathcal{K}''$. 
Evaluating Eq. \eqref{eq:lambdappleft} at $\chi=0$ gives
\begin{equation}
      \langle \lambda''| \boldsymbol{R} +2\langle \lambda'| \dot { \boldsymbol{H} }  = \lambda''\langle \lambda|+2\lambda'\langle \lambda'|.  
\end{equation}
Multiplying both sides by $\boldsymbol{R}^+$ we have
\begin{equation}
    \langle \lambda''| (\boldsymbol{I}- |\lambda\rangle \langle\lambda|) +2\langle \lambda'| \dot { \boldsymbol{H} }\boldsymbol{R}^+  = 2\lambda'\langle \lambda'|\boldsymbol{R}^+,
\end{equation}
and by intermediate normalization $\langle \lambda''|\lambda \rangle =0$, then $\langle \lambda'' |$ is given by
\begin{equation}
    \langle \lambda''| =2\lambda'\langle \lambda'|\boldsymbol{R}^+ -2\langle \lambda'| \dot { \boldsymbol{H} }\boldsymbol{R}^+ .
\end{equation}
Recall that $\lambda', |\lambda'\rangle$ an $ \langle \lambda'|$ are all on the order of $\dot {\boldsymbol{H}} = {O} ( \tau^{-1})$, then $\langle \lambda''| = O(\tau^{-2})$.
Similar argument applies to $|\lambda '' \rangle$, too.
Taking second derivatives of the last two terms (geometric phase and boundary terms) in Eq. \eqref{eq:Kchi} with respect to $i\chi$ and evaluating at $\chi=0$, one can see that all terms are of order $O(\tau^{-2})$.



{
%
}

\end{document}